\documentclass[
superscriptaddress,
preprintnumbers,
twocolumn,
nofootinbib,
amsmath,amssymb,
aps,
]{revtex4-2}

\usepackage{graphicx}
\usepackage{dcolumn}
\usepackage{bm}
\usepackage{hyperref}

\usepackage{comment}
\usepackage{caption}
\usepackage{subcaption}
\usepackage{booktabs}
\numberwithin{equation}{section}
\usepackage{color}
\usepackage{mathtools}
\usepackage[normalem]{ulem} 
\usepackage{amsmath} 
\usepackage{orcidlink}

\usepackage{url}
\hypersetup{colorlinks=true, citecolor=cyan}

\newcommand{\be}{\begin{equation}}
\newcommand{\ee}{\end{equation}}

\definecolor{mygreen}{RGB}{0,130,0} 

\begin{document}
\preprint{RESCEU-20/25}

\author{Daiki Watarai\orcidlink{0009-0002-7569-5823}}
\affiliation{Department of Physics, Graduate School of Science, The University of Tokyo, 7-3-1 Hongo, Bunkyo-ku, Tokyo 113-8655, Japan}
\affiliation{Research Center for the Early Universe (RESCEU), Graduate School of Science, The University of Tokyo, Tokyo 113-0033, Japan}

\author{Atsushi Nishizawa \orcidlink{0000-0003-3562-0990}}
\affiliation{Physics Program, Graduate School of Advanced Science and Engineering, Hiroshima University, Higashi-Hiroshima, Hiroshima 739-8526, Japan}
\affiliation{Astrophysical Science Center, Hiroshima University, Higashi-Hiroshima, Hiroshima 739-8526, Japan}

\author{Hiroki Takeda \orcidlink{0000-0001-9937-2557}}
\affiliation{The Hakubi Center for Advanced Research, Kyoto University, Kyoto 606-8501, Japan}
\affiliation{Department of Physics, Kyoto University, Kyoto 606-8502, Japan}

\author{Hayato Imafuku \orcidlink{0009-0001-3490-8063}} 
\affiliation{Department of Physics, Graduate School of Science, The University of Tokyo, 7-3-1 Hongo, Bunkyo-ku, Tokyo 113-8655, Japan}
\affiliation{Research Center for the Early Universe (RESCEU), Graduate School of Science, The University of Tokyo, Tokyo 113-0033, Japan}

\author{Kipp Cannon \orcidlink{0000-0003-4068-6572}}
\affiliation{Research Center for the Early Universe (RESCEU), Graduate School of Science, The University of Tokyo, Tokyo 113-0033, Japan}

\newcommand*{\diff}{\,\mathrm{d}}
\newcommand*{\g}{\ensuremath{\hat{\gamma}_1}}
\newcommand*{\ppca}{\ensuremath{\hat{P}_\mathrm{PCA}}}
\newcommand*{\ppds}{posterior probability distributions~}
\newcommand*{\Ppds}{Posterior probability distributions~}

\date{\today}

\begin{abstract}
Gravitational waves from compact binary coalescences provide unique opportunities to test general relativity (GR) in the strong-field regime. 
In particular, the merger phase, during which two compact objects finally coalesce, corresponds to the regime of the strongest gravitational fields accessible by direct observation and thus serves as a probe of the nonlinear nature of gravity. 
In this work, we test GR in the merger phase by analyzing GW150914 using a modified waveform proposed in [Watarai \textit{et al.} 2024], which parametrizes possible deviations from GR during this stage. 
Within this framework, the inferred deviation parameters can be translated into model-independent constraints on physically meaningful quantities. 
For GW150914, we find that the additional energy radiated in the merger phase is constrained to be $0.26^{+0.75}_{-0.62}~\%$ of the total energy emitted over the entire coalescence predicted by GR, and the deviation in the coalescence time is $2.17^{+9.56}_{-9.90}~\mathrm{ms}$, both within the $90\%$ credible interval. 
These two constraints serve as observational benchmarks for deviations in the nonlinear gravity regime, offering guidance for theoretical investigations of beyond-GR models.
\end{abstract}

    \title{Observational constraints on the nonlinear regime of gravity with a parametrized beyond-GR gravitational waveform model}

\maketitle
\footnotetext[1]{Email: \href{mailto:your.name@example.com}{wataraidaiki@resceu.s.u-tokyo.ac.jp}}

\section{Introduction}
General relativity (GR), the standard theory of gravity, has passed all experimental and observational tests to date, particularly in weak-field regimes~\cite{Will:2014kxa, Berti:2015itd}. 
However, from a theoretical standpoint, GR is not regarded as the ultimate theory of gravity, due to issues such as non-renormalizability as a quantum theory and the prediction of singularities where the classical description breaks down. 
To explore physics in strong gravity, it is essential to test GR in the strong-field, highly dynamical regime, where deviations might arise.

Gravitational-wave (GW) astronomy now provides a unique opportunity to probe this regime observationally. 
In the past decade, the detection of GWs from compact binary coalescences (CBCs) has enabled us to study gravity in previously inaccessible regimes~\cite{LIGOScientific:2016aoc, LIGOScientific:2016vlm}. 
The LIGO–Virgo–KAGRA (LVK) collaboration has reported hundreds of GW events~\cite{LIGOScientific:2018mvr, LIGOScientific:2020ibl, KAGRA:2021vkt, LIGOScientific:2025slb} and conducted various tests of GR, so far finding no statistically significant deviations within current sensitivities~\cite{LIGOScientific:2016vlm, LIGOScientific:2019fpa, LIGOScientific:2020tif, LIGOScientific:2021sio, KAGRA:2025oiz, LIGOScientific:2025obp}.

A binary black hole (BBH) coalescence is typically divided into three phases: inspiral, merger, and ringdown. 
The inspiral and ringdown phases can be modeled using perturbative methods, post-Newtonian (PN) theory for the inspiral and BH perturbation theory for the ringdown. 
In contrast, the merger phase involves fully nonlinear dynamics for which no analytical methods are established, and numerical relativity (NR) simulations are required to understand the system.

Parameterized frameworks have been widely developed for the inspiral and ringdown phases of compact binary coalescences, enabling systematic tests of GR in those regimes~(e.g.~\cite{Yunes:2009ke, Mishra:2010tp, Li:2011cg, Saleem:2021nsb, Mehta:2022pcn, Gossan:2011ha, Meidam:2014jpa, Glampedakis:2017dvb, Brito:2018rfr, Cardoso:2019mqo, McManus:2019ulj, Carullo:2019flw, Ghosh:2021mrv, Ghosh:2024het, Berti:2025hly, KAGRA:2025oiz, LIGOScientific:2025obp, Others:2025nbi}). 
In contrast, constructing a comparable framework for the merger phase that allows direct comparison with beyond-GR theories remains challenging. 
Nevertheless, phenomenological parameterizations of deviations in the merger phase have been explored in several studies~\cite{Bonilla:2022dyt, Maggio:2022hre, Pompili:2025cdc}. 
For instance, Maggio \textit{et al.} and Pompili \textit{et al.} constrained deviations in the fundamental QNM frequency from real data, with the former also estimating an amplitude shift in the nonlinear regime.

The LVK collaboration has likewise constrained fractional deviations in phenomenological coefficients of NR-calibrated inspiral–merger–ringdown waveform models~\cite{LIGOScientific:2016aoc, LIGOScientific:2020tif}. 
While informative, these analyses remain tied to specific waveform models, and correlations between GR and deviation parameters can enlarge statistical errors and introduce possible systematic biases. 
This motivates the development of alternative approaches that can provide robust, physically interpretable, and model-independent constraints in the merger regime.

To address the issue above, in our previous work~\cite{Watarai:2023yky} we proposed a phenomenological modification of the \texttt{IMRPhenomD} waveform~\cite{Husa:2015iqa, Khan:2015jqa}, introducing two additional beyond-GR parameters—one in the amplitude and one in the phase—constructed to remain consistent with radiation-reaction effects. 
This framework offers two key advantages compared with existing works. 
First, the beyond-GR parameters can be reinterpreted as model-independent observational constraints on physically meaningful quantities, such as the excess energy radiated during the merger and the shift in the coalescence time. 
Second, the parametrization is designed to reduce parameter correlations with the standard GR parameters, thereby enabling clearer inference of possible deviations. 
Our model-independent constraints presented in this work will provide concrete benchmarks that theoretical studies—including numerical simulations in modified gravity~(e.g.~\cite{Okounkova:2017yby, Okounkova:2020rqw, East:2020hgw, Figueras:2021abd, Corman:2022xqg, Cayuso:2023xbc, Corman:2024cdr})—can investigate against, thereby strengthening the connection between data analysis and theory.

We apply our modified waveform to GW data and constrain possible deviations from GR in the nonlinear regime of gravity.
We first carry out two sets of injection–recovery studies using both GR and non-GR injections to evaluate the robustness of the method against false detections of deviations and its ability to recover beyond-GR effects. 
We then analyze GW150914~\cite{LIGOScientific:2016aoc}, obtaining model-independent constraints on the additional energy radiated during the merger phase and on the deviation in the coalescence time.

This paper is organized as follows. 
In Sec.~\ref{sec:modified_waveform}, we summarize the modified waveform for capturing beyond-GR effects in the merger phase of BBH coalescences, which is proposed in Ref.~\cite{Watarai:2023yky}. 
In Sec.~\ref{sec:injection_recovery}, we present injection recovery tests using the modified waveform. 
We perform two types of injections: GR and non-GR, to assess whether the modified waveform produces apparent deviations in GR cases and whether it can correctly recover deviations in non-GR cases. 
Parameter estimation results for GW150914 are reported in Sec.~\ref{sec:analysis_of_real_data}. 
In Sec.~\ref{sec:discussion}, we discuss our findings in the context of previous works and outline possible extensions of our method, including connections with LVK analyses and current limitations. 
We finally conclude in Sec.~\ref{sec:conclusion} by summarizing the main results and highlighting directions for future research.

\section{Modified Waveform}
\label{sec:modified_waveform}
In this section, we review the modified waveform introduced in Ref.~\cite{Watarai:2023yky}, which is employed in the subsequent analysis. This waveform is built based on \texttt{IMRPhenomD}~\cite{Khan:2015jqa} and extends it by introducing two beyond-GR parameters: $\g$, which controls the overall amplification in the merger phase and consequently causes additional losses of energy and angular momentum radiation, and $\ppca$, which is obtained from principal component analysis (PCA)\footnote{Application of PCA to GW data analysis has been done in several works~(e.g.~\cite{Saleem:2021nsb, Datta:2022izc, Seymour:2024kcd, Mahapatra:2025cwk}). See also Refs.~\cite{Shimomura:2025fhz, Kume:2025wpl} for the application of independent component analysis.} of the phase parameters in the nonlinear regime and related to peak-time and phase shifts.

In Sec.~\ref{sec:PhenomD}, we outline the structure of \texttt{IMRPhenomD}, which serves as the foundation of our model. In section~\ref{sec:sub_modified_waveform}, we elaborate on the modified waveform.

\subsection{IMRPhenomD}
\label{sec:PhenomD}
\texttt{IMRPhenomD} describes the dominant $(l,m)=(2,\pm 2)$ mode from the whole stages of an aligned-spin BBH coalescence in the frequency domain, with the mass ratio up to $1:18$ \cite{Khan:2015jqa}. The form is given by
\begin{equation}
\label{eq:IMRPhenomD}
    \tilde{h}_\mathrm{GR}(f) = A_\mathrm{GR}(f)\:\mathrm{e}^{-\mathrm{i}\phi_\mathrm{GR}(f)}\;,
\end{equation}
where $\phi_\mathrm{GR}(f)$ and $A_\mathrm{GR}(f)$ are the GW phase and amplitude. In the following, we fix a subscript ``GR'' to a quantity in GR. $\phi_\mathrm{GR}(f)$ and $A_\mathrm{GR}(f)$ are divided into three frequency ranges: the inspiral (ins), intermediate (int), and merger-ringdown (MR), respectively,
\begin{gather}
\label{PhenomD_phi}
\phi_\mathrm{GR}(f) = \left\{
\begin{array}{lll}
\phi_{\mathrm{ins}}(f), & f \leq f_{\mathrm{p1}}\\
\phi_{\mathrm{int}}(f), & f_{\mathrm{p1}} \leq f \leq f_{\mathrm{p2}}\\
\phi_{\mathrm{MR}}(f), & f \geq f_{\mathrm{p2}}
\end{array}\;,
\right.\\
\label{PhenomD_A}
A_\mathrm{GR}(f) = \left\{
\begin{array}{lll}
A_{\mathrm{ins}}(f), & f \leq f_{\mathrm{a1}}\\
A_{\mathrm{int}}(f), & f_{\mathrm{a1}} \leq f \leq f_{\mathrm{a2}}\\
A_{\mathrm{MR}}(f), & f \geq f_{\mathrm{a2}} 
\end{array}\;,
\right.
\end{gather}
where the explicit forms of the functions and the collocation frequencies $f_{\mathrm{p}i, \mathrm{a}i}$ ($i=1,2$) are shown in Ref.~\cite{Khan:2015jqa}. 
The functions are constructed in a way that reflects the physics of each regime. 
The inspiral part is based on the PN expansions supplemented with calibration terms to improve accuracy. 
The intermediate regime is described by phenomenological polynomial fits that ensure smooth transitions between the inspiral and merger–ringdown parts. 
Finally, the merger–ringdown part is modeled with damped sinusoidal functions capturing the QNM behavior of the remnant BH, with coefficients calibrated against numerical relativity simulations.

\subsection{Modified waveform}
\label{sec:sub_modified_waveform}
The modified waveform is expressed as
\begin{equation}
\label{eq:modified_waveform_final}
\tilde{h}_\mathrm{MG}(f;\g,\ppca) := A_\mathrm{MG}(f;\g)\,\mathrm{e}^{-\mathrm{i}\phi_\mathrm{MG}(f;\g,\ppca)}\;,
\end{equation}
where $A_\mathrm{MG}(f;\g)$ and $\phi_\mathrm{MG}(f;\g,\ppca)$ denote the modified amplitude and phase in the frequency domain. The following sections outline $A_\mathrm{MG}(f;\g)$ and $\phi_\mathrm{MG}(f;\g,\ppca)$; explicit functional forms are provided in Sec.~III of Ref.~\cite{Watarai:2023yky}.

The parameter $\g$ controls an overall amplification from the GR waveform in the merger phase, corresponding to additional energy and angular momentum radiation relative to GR predictions.  
On the other hand, $\ppca$ characterizes peak-time and phase shifts, obtained through PCA of the fractional deviations in the phenomenological phase parameters of \texttt{IMRPhenomD} in the nonlinear gravity regime.

Since the amplitude and phase in \texttt{IMRPhenomD} are modeled separately, our two beyond-GR parameters are likewise introduced independently. 
In reality, a specific modified gravity theory would relate deviations in the amplitude to corresponding shifts in the GW phase through modifications of the binary dynamics up to the merger phase. 
Here, however, we treat the amplitude and phase deviations as independent phenomenological parameters in the waveform construction.

\subsubsection{\textbf{Physical consistency}}
\label{sec:physical_consistency}
The amplification introduced by $\g$ leads to additional energy and angular momentum radiation. 
Consequently, the remnant mass and spin must be corrected to account for these losses.
We incorporate the corresponding radiation–reaction effects into the waveform.  

For a given value of $\g$, the final mass and spin are expressed as
\begin{align}
    \bar{M}_\mathrm{f} &= M_\mathrm{ini} - \Delta E\:,\\
    \bar{J}_\mathrm{f} &= J_\mathrm{ini} - \Delta J\:,
\end{align}
where $M_\mathrm{ini}$ and $J_\mathrm{ini}$ denote the total mass and angular momentum of the binary system, and $\Delta E$ and $\Delta J$ are the total radiated energy and angular momentum over the entire coalescence process.  
These quantities can be decomposed into the GR contributions, $\Delta E_\mathrm{GR}$ and $\Delta J_\mathrm{GR}$, and the additional radiation due to beyond-GR effects:
\begin{align}
    \Delta E &= \Delta E_\mathrm{GR} + \delta\Delta E(\g; M, \eta, \chi_\mathrm{eff})\:,\\
    \Delta J &= \Delta J_\mathrm{GR} + \delta\Delta J(\g; M, \eta, \chi_\mathrm{eff})\:,
\end{align}
where $\delta\Delta E(\g; M, \eta, \chi_\mathrm{eff})$ and $\delta\Delta J(\g; M, \eta, \chi_\mathrm{eff})$ are the additional radiated energy and angular momentum induced by $\g$, computed using the GR radiation formula (see Eqs.~(III.33) and (III.34) in Ref.~\cite{Watarai:2023yky}).  
This procedure modifies the QNM frequency in the MR regime.  

In generating an IMR waveform, the key quantities are the remnant mass $\bar{M}_\mathrm{f}$ and the dimensionless spin parameter
\begin{equation}
    \bar{a}_\mathrm{f} := \frac{\bar{J}_\mathrm{f}}{{\bar{M}_\mathrm{f}}^2}\:.
\end{equation}
For efficient waveform generation in data analysis, we provide fitting formulas for $\bar{M}_\mathrm{f}$ and $\bar{a}_\mathrm{f}$ in Appendix~\ref{app:fitting_formula} to reduce the computational cost in parameter estimation.

\subsubsection{\textbf{Modified GW amplitude}}
\label{sec:modified_amp}
The modified amplitude $A_\mathrm{MG}(f;\g)$ is defined as
\begin{equation}
\label{eq:A_MR}
A_\mathrm{MG}(f;\g) :=
\begin{cases}
A_{\mathrm{ins}}(f), & f \leq f_{\mathrm{a1}},\\
\bar{A}_{\mathrm{int\_m}}(f;\g), & f_{\mathrm{a1}} \leq f \leq \bar{f}_{\mathrm{a2}},\\
\bar{A}_{\mathrm{MR\_m}}(f;\g), & f \geq \bar{f}_{\mathrm{a2}},
\end{cases}
\end{equation}
where the subscript ``m'' denotes a function modified from \texttt{IMRPhenomD}, and an overbar indicates the inclusion of radiation–reaction effects due to $\g$ (see Sec.~\ref{sec:physical_consistency}).  

The parameter $\g$ describes deviation from GR in the nonlinear regime. 
Starting from $A_\mathrm{int}(f)$ and $A_\mathrm{MR}(f)$ in the GR waveform, we introduce $\g$ such that
\begin{equation}
\label{eq:g1_MR}
\begin{split}
    A_{\mathrm{MR\_m}}(f;\g) &:= (1+\g)A_\mathrm{MR}(f) \\
    &=: A_\mathrm{MR}(f) + \Delta A_\mathrm{MR}(f;\g)\:,
\end{split}
\end{equation}
which amplifies the peak amplitude.  
This amplification is then smoothly interpolated into the intermediate regime by modifying the phenomenological coefficients of $A_\mathrm{int}(f)$ at the matching frequency $f_{\mathrm{a2}}$.

\subsubsection{\textbf{Modified GW phase}}
The modified phase ${\phi}_\mathrm{MG}(f;\g, \ppca)$ is defined as
\begin{equation}
\label{eq:phi_MR}
\begin{split}
\phi_\mathrm{MG}(f;\g, \ppca) :=
\begin{cases}
\phi_{\mathrm{ins}}(f), & f \leq f_{\mathrm{p1}},\\
\bar{\phi}_{\mathrm{int\_m}}(f;\hat{P}_{\mathrm{PCA}}), & f_{\mathrm{p1}} \leq f \leq \bar{f}_{\mathrm{p2}},\\
\bar{\phi}_{\mathrm{MR\_m}}(f;\g, \hat{P}_{\mathrm{PCA}}), & f \geq \bar{f}_{\mathrm{p2}},
\end{cases}
\end{split}
\end{equation}
where $\ppca$ is the first principal component of the fractional deviations in the phenomenological phase parameters $\{\beta_2, \beta_3, \alpha_2, \alpha_3\}$ of \texttt{IMRPhenomD}.  
Specifically, denoting the fractional errors by $\{ \hat{\Lambda}_i \}:=\{ \hat{\beta}_2, \hat{\beta}_3, \hat{\alpha}_2, \hat{\alpha}_3 \}$, $\ppca$ is defined by the first principal component of the Fisher matrix,
\begin{equation}
\label{eq:Fisher_matrix}
    \boldsymbol{F}_{\hat{\beta}\hat{\alpha}} = 
    \begin{pmatrix}
    F_{\hat{\beta}_2\hat{\beta}_2} & F_{\hat{\beta}_2\hat{\beta}_3} & F_{\hat{\beta}_2\hat{\alpha}_2} & F_{\hat{\beta}_2\hat{\alpha}_3} \\
     & F_{\hat{\beta}_3\hat{\beta}_3} & F_{\hat{\beta}_3\hat{\alpha}_2} & F_{\hat{\beta}_3\hat{\alpha}_3} \\
     &  &
    F_{\hat{\alpha}_2\hat{\alpha}_2} & F_{\hat{\alpha}_2\hat{\alpha}_3} \\
    \mathrm{sym.} &  &
     & F_{\hat{\alpha}_3\hat{\alpha}_3}
    \end{pmatrix}\;,
\end{equation}
with
\begin{equation}
    \left(\boldsymbol{F}_{\hat{\beta}\hat{\alpha}}\right)_{ij} := 
    \int_{f_\mathrm{min}}^{f_\mathrm{max}} \frac{\partial_{\hat{\lambda}_i} \tilde{h}(f) \:\partial_{\hat{\lambda}_j} \tilde{h}(f)}{S_\mathrm{n}(f)}\:\diff f
\end{equation}
where $\partial_{\hat{\Lambda}_i}$ indicates the partial derivative with respect to $\hat{\Lambda}_i$, $S_\mathrm{n}(f)$ is the power spectral density of detectors, but set to $S_\mathrm{n}(f)=1\:\mathrm{Hz}^{-1}$ here, $f_\mathrm{min}$ is set to $0.0035/M$, which is the minimum frequency considered in \texttt{IMRPhenomD}, and $f_\mathrm{max}$ is the upper cutoff frequency, chosen sufficiently large to ensure convergence.

Nonzero values of $\ppca$ shift the peak time and phase with a certain combination depending on the masses and spins of a system.  
The analytic expressions derived in Appendix~\ref{app:formulas_shifts} are obtained under the assumption that the reaction of $\g$ on the remnant mass and spin is neglected.

\subsection{Time-domain modified waveform}
\begin{figure*}
    \centering
    \includegraphics[width=\linewidth]{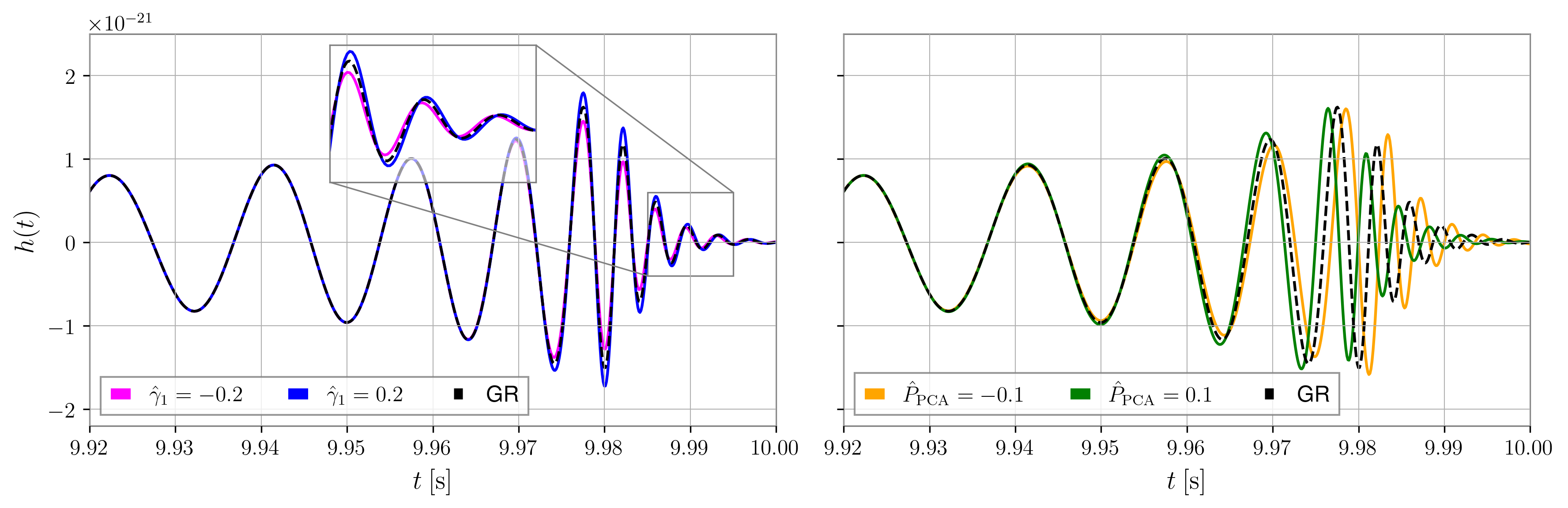} 
    \caption{
    Left: Comparison of waveforms with $\g = -0.2$ (magenta), $\g = 0.2$ (blue), and the GR case (black dotted).  
    Since $\g$ induces additional energy and angular momentum radiation, the remnant mass and spin are modified from the GR prediction, resulting in a change in the QNM frequency.  
    Right: Comparison of waveforms with $\ppca = -0.1$ (orange), $\ppca = 0.1$ (green), and the GR case (black dotted).  
    Non-zero values of $\ppca$ introduce phase and time shifts relative to the GR waveform.
    }\label{fig:modified_waveform}
\end{figure*}

Figure~\ref{fig:modified_waveform} presents the modified and GR waveforms in the time domain, assuming equal-mass binaries with a total mass of $M=60M_\odot$ and the effective spin parameter of $\chi_\mathrm{eff}=0$.
The left panel shows waveforms with $\g = -0.2$ (magenta) and $\g = 0.2$ (blue), compared with the GR case (black dotted). 
Since $\g$ alters the amount of radiated energy and angular momentum, it changes the remnant properties and hence the QNM frequency. 
The right panel shows waveforms with $\ppca = -0.1$ (orange) and $\ppca = 0.1$ (green), again compared with the GR case. 
Nonzero values of $\ppca$ introduce relative shifts in the peak time and phase with respect to the GR waveform.

\begin{table*}
\caption{Injection cases considered in this study. 
The table lists the source parameters for GR and non-GR injections, including component source-frame masses $m_{1,2}$, spin parameters $\chi_\mathrm{eff}:=(m_1\chi_1+m_2\chi_2)/(m_1+m_2)$ and $\delta \chi:=\chi_1-\chi_2$, beyond-GR parameters, and extrinsic parameters. 
The network SNR values correspond to the design sensitivities of LIGO Hanford, LIGO Livingston, Virgo, and KAGRA.}
\label{tb:inj_source_GR}
\begin{ruledtabular}
\begin{tabular}{ccccccccccccccc}
 Type & $m_1\:[M_\odot]$ & $q:=\frac{m_2}{m_1}$ & $\chi_{\mathrm{eff}}$ & $\delta \chi$ & $\g$ & $\ppca$ & $d_\mathrm{L}$ [Mpc] 
 & $t_c$ [s] & $\phi_\mathrm{c}$ & $\alpha$ & $\delta$ & $\iota$ & $\psi$ & $\mathrm{SNR}_\mathrm{net}$  \\ \hline
    GR & $30$ & $1$ &  $\{ -0.5,\:0,\:0.5\}$ & $0$ & $0$ & $0$ & $400$ & $0$ & $0$ & $0$ & $0$ & $0$ & $0$ & $\{99.7, 113, 129\}$ \\
    ${}$ & $10$ & $1$ &  $0$ & $0$ & $0$ & $0$ & $400$ & $0$ & $0$ & $0$ & $0$ & $0$ & $0$ & $45.8$ \\ \colrule
    Non-GR & $30$ & $1$ &  $0$ & $0$ & $\{-0.2, 0.2\}$ & $0$ & $400$ & $0$ & $0$ & $0$ & $0$ & $0$ & $0$ & $\{108, 118\}$ \\
    ${}$ & $30$ & $1$ &  $0$ & $0$ & $0$ & $\{-0.1, 0.1\}$ & $400$ & $0$ & $0$ & $0$ & $0$ & $0$ & $0$ & $\{113, 113\}$ \\
\end{tabular}
\end{ruledtabular}
\end{table*}

\begin{table}
\caption{\label{tb:priors}%
Priors for the analysis parameters. 
Here, $\mathcal{M}_z := \eta^{3/5}(m_{1\_z}+m_{2\_z})$ denotes the redshifted chirp mass. 
For the injection with $m_1 = 10~M_\odot$ and for GW190728 (Appendix~\ref{app:GW190728}), the lower bound for $\mathcal{M}_\mathrm{c}$ is set to $5~M_\odot$.}
\begin{ruledtabular}
\begin{tabular}{ccc}
Parameter & Range & Prior \\ \hline
    $m_{1\_z,2\_z}\:[M_\odot]$ & $[5,\:80]$ & uniform \\
    $q:=m_2/m_1$ & [0.125,\:1] & set the bounds \\
    $\mathcal{M}_{\mathrm{c}\_z}\:[M_\odot]$ & [20(5),\:100] & set the bounds \\
    $\chi_{1,2}$ & $[-0.99,\:0.99]$ & uniform \\
    $d_\mathrm{L}\:[\mathrm{Mpc}]$ & $[50,\:2000]$ & uniform in volume \\
    $t_\mathrm{c}$ [s] & [-0.1+$t_\mathrm{peak}$, 0.1+$t_\mathrm{peak}$] & uniform  \\
    $\phi_\mathrm{c}$ & [$0,\:2\pi$] & uniform \\
    $\alpha$ & [$0,\:2\pi$] & uniform \\
    $\delta$ & [$0,\:\pi$] & uniform in $\cos\delta$ \\
    $\iota$ & [$0,\:\pi$] & uniform in $\sin\iota$ \\
    $\psi$ & [$0,\:2\pi$] & uniform \\
    $\g$ & [$-5, 5$] & uniform \\
    $\ppca$ & [$-5, 5$] & uniform \\
\end{tabular}
\end{ruledtabular}
\end{table} 

\begin{figure*}
    \centering
    \includegraphics[width=0.8\linewidth]{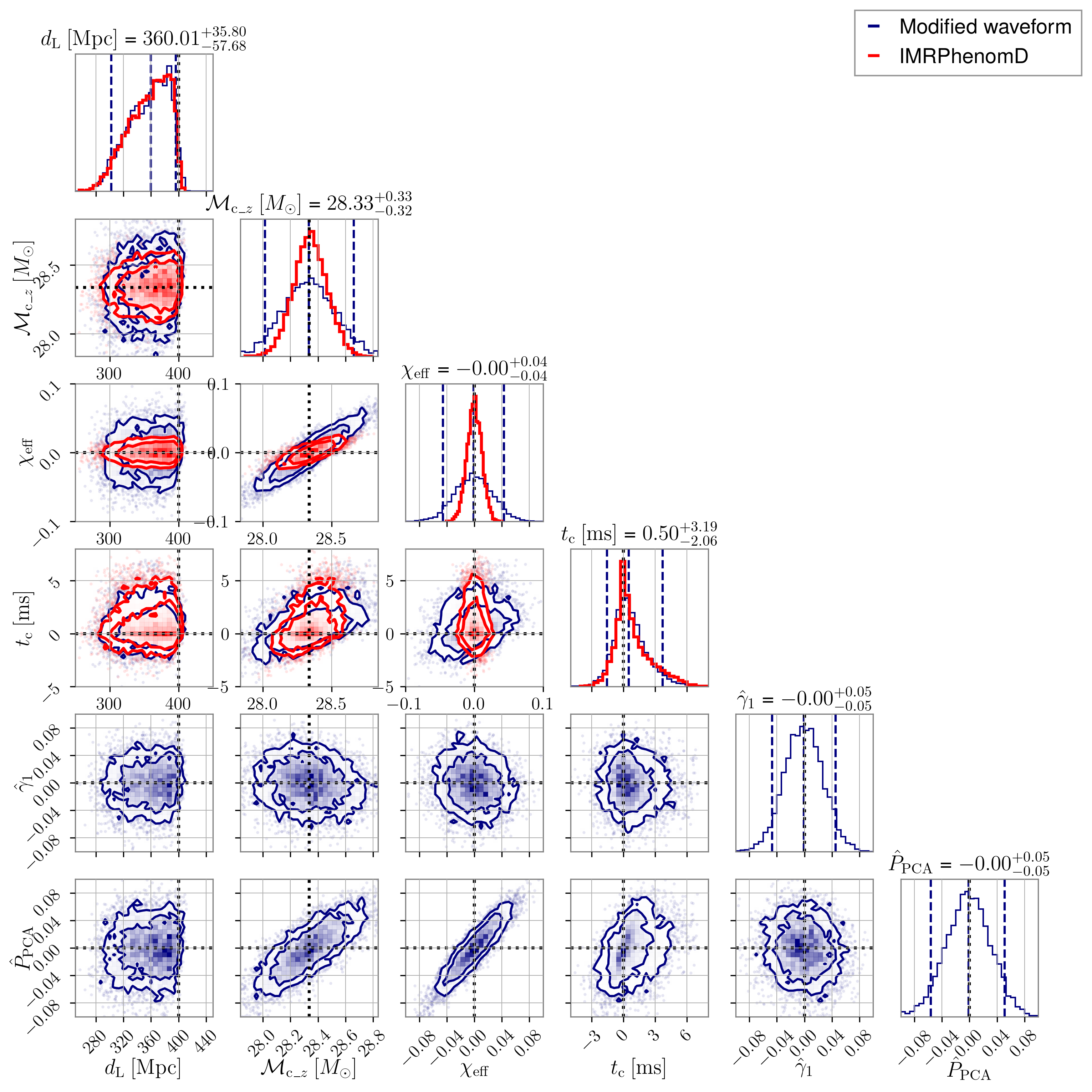}
    \caption{\Ppds for $\{ d_\mathrm{L}, \mathcal{M}_{c\_z}, \chi_\mathrm{eff}, t_\mathrm{c}, \g, \ppca\}$ with a GR injection. The navy and red show the recovery with the modified waveform and the GR waveform. The inner and outer contours indicate $68\%$ and $90\%$ CIs. Black dotted and navy dashed lines show the fiducial values, mean value, and $90\%$ CI, respectively.}
    \label{fig:GR_inj}
\end{figure*}

\section{Injection recovery}
\label{sec:injection_recovery}
Before analyzing observational data, we perform a series of injection studies to validate our method. These studies address two key questions:  
\begin{itemize}
    \item[(i)] whether the modified waveform recovers deviation parameters consistent with zero when the injected signal is a GR waveform.
    \item[(ii)] whether the modified waveform can accurately recover nonzero deviations when they are present in the signal.
\end{itemize}

We introduce the Bayesian data analysis framework in Sec.~\ref{sec:bayesian} and the specific injection cases in Sec.~\ref{sec:cases}. We present the results for GR injections in Sec.~\ref{sec:GR_injection} and for non-GR injections in Sec.~\ref{sec:non-GR_injection}, corresponding to questions (i) and (ii), respectively.

\subsection{Bayesian analysis framework}
\label{sec:bayesian}
Our inference is based on Bayesian analysis, which computes the posterior probability distribution $p(\vec{\theta}|\vec{d},M)$ according to Bayes' theorem:
\begin{equation}
    p(\vec{\theta}|\vec{d},M) = \frac{p(\vec{\theta}|M)\, p(\vec{d}|\vec{\theta},M)}{p(\vec{d}|M)}\:,
\end{equation}
where $\vec{\theta}$ is the set of model parameters, $\vec{d}$ denotes the data observed by the detectors, $M$ is the waveform model, $p(\vec{\theta}|M)$ is the prior distribution, $p(\vec{d}|\vec{\theta},M)$ is the likelihood function, and $p(\vec{d}|M)$ is the evidence.  

The model parameters for the GR and modified waveforms are defined as
\begin{align}
\label{eq:model_parameters}
    \vec{\theta}_\mathrm{GR} &= \{ m_{1\_z}, m_{2\_z}, \chi_1, \chi_2, d_\mathrm{L}, t_\mathrm{c}, \phi_\mathrm{c}, \alpha, \delta, \iota, \psi \}\:,\\
    \vec{\theta}_\mathrm{MG} &= \vec{\theta}_\mathrm{GR} \cup \{ \g, \ppca \}\:,
\end{align}
where $m_{i\_z}$ and $\chi_{i}$ ($i=1,2$) are the component redshifted masses and dimensionless spins, $d_\mathrm{L}$ is the luminosity distance, $t_\mathrm{c}$ is the coalescence time, $\phi_\mathrm{c}$ is the coalescence phase, $\alpha$ is the right ascension, $\delta$ is the declination, $\iota$ is the inclination angle, and $\psi$ is the polarization angle. Throughout this analysis, we assume the flat $\Lambda$ cold dark matter cosmology with parameters from Ref.~\cite{Planck:2015fie}.  

We inject signals with zero noise into the outputs from the LIGO Hanford, LIGO Livingston, Virgo, and KAGRA detectors, assuming their design sensitivities~\cite{LIGOScientific:2014pky, VIRGO:2014yos, KAGRA:2020tym}. The prior distributions are summarized in Table~\ref{tb:priors}. For $\vec{\theta}_\mathrm{GR}$, we largely follow the setup of Ref.~\cite{LIGOScientific:2016vlm}, and we adopt uniform priors in $[-5, 5]$ for $\g$ and $\ppca$. 
Note that these ranges exceed the validity regime of $\ppca$ and $\g$, as they are assumed to be small enough in the waveform construction (see Sec.~III.~F of Ref.~\cite{Watarai:2023yky}). 
The ranges are chosen to be large enough for the analysis.
The Bayesian analysis is performed using the \texttt{Bilby} software package~\cite{Ashton:2018jfp} with the \texttt{Dynesty} sampler~\cite{2020MNRAS.493.3132S}. 
We use $n_\mathrm{live}=1000$ for most cases, and increase it to $1500$ for the smaller-mass injections and for GW190728 in Appendix~\ref{app:GW190728}. 
This choice was confirmed to be sufficient to ensure convergence of the sampling procedure.

\subsection{Injected signals}
\label{sec:cases}

We consider two types of signal injections: GR and non-GR, as summarized in Table~\ref{tb:inj_source_GR}.  
The quantities with subscript $z$ denote detector-frame values.

For the GR injections presented in Sec.~\ref{sec:GR_injection}, we use equal-mass binaries with total masses of $M = 20~M_\odot$ and $60~M_\odot$ and effective spins of $\chi_\mathrm{eff} = -0.5, 0,$ and $0.5$. 
The equal-mass configuration is chosen since the $(2,2)$ mode is expected to be dominant, allowing us to focus on the main contribution to the signal.  
These cases extend the parameter sets analyzed in the Fisher study of Ref.~\cite{Watarai:2023yky}, where nonspinning systems were considered, by additionally exploring the impact of aligned spins.  
In other words, we analyze not only the same parameter sets as Ref.~\cite{Watarai:2023yky} but also additional cases with varying spins.

For the non-GR injections, we vary $\g$ and $\ppca$ while fixing the component masses to $m_{1,2} = 30~M_\odot$ and $\chi_\mathrm{eff} = 0$. Here, we focus on the heavier BBH system with no spins studied in Ref.~\cite{Watarai:2023yky}, which was found to provide stronger sensitivity to beyond-GR effects in the Fisher analysis than the smaller-mass case. 
The values of the beyond-GR parameters $\g$ and $\ppca$ are chosen within the validity range of the model.

\subsection{GR injections}
\label{sec:GR_injection}
We perform GR injections using the parameter sets listed in Table~\ref{tb:inj_source_GR}.  
Figure~\ref{fig:GR_inj} shows the \ppds for  
$\{ d_\mathrm{L}, \mathcal{M}_{c\_z}, \chi_\mathrm{eff}, t_\mathrm{c}, \g, \ppca \}$  
for the injection corresponding to the first row of Table~\ref{tb:inj_source_GR} with $\chi_\mathrm{eff}=0$.  
Results obtained with the modified waveform and \texttt{IMRPhenomD} are shown in navy and red, respectively.  
The inner and outer contours represent the $68\%$ and $90\%$ credible intervals (CIs).
The black dotted lines denote the fiducial values, while the dashed lines indicate the posterior means and 90\% CIs in the diagonal plots. 
These plotting conventions are used throughout this paper.

For the GR parameters $\{ d_\mathrm{L}, \mathcal{M}_{c\_z}, \chi_\mathrm{eff}, t_\mathrm{c} \}$,  
the posteriors obtained with the modified waveform agree well with those from \texttt{IMRPhenomD}. 
The existence of parameter correlations can sometimes cause parameter estimation bias when specific events in the parameter space are selected (e.g.,~\cite{Usman:2018imj, Imafuku:2025fta}).
Our result demonstrates that the inclusion of $\g$ and $\ppca$ does not significantly bias the estimation of $\vec{\theta}_\mathrm{GR}$ through parameter correlations. 
Moreover, the recovered values of $\g$ and $\ppca$ are consistent with zero within $68\%$ CIs, indicating that the modified waveform does not produce spurious deviations from GR.  
At the same time, it is found that $\ppca$ is modestly correlated with $\mathcal{M}_{c\_z}$ and $\chi_\mathrm{eff}$, which leads to somewhat larger uncertainties in these parameters.
The results for the other GR injection cases are qualitatively similar.

\begin{figure}
    \centering
    \includegraphics[width=\linewidth]{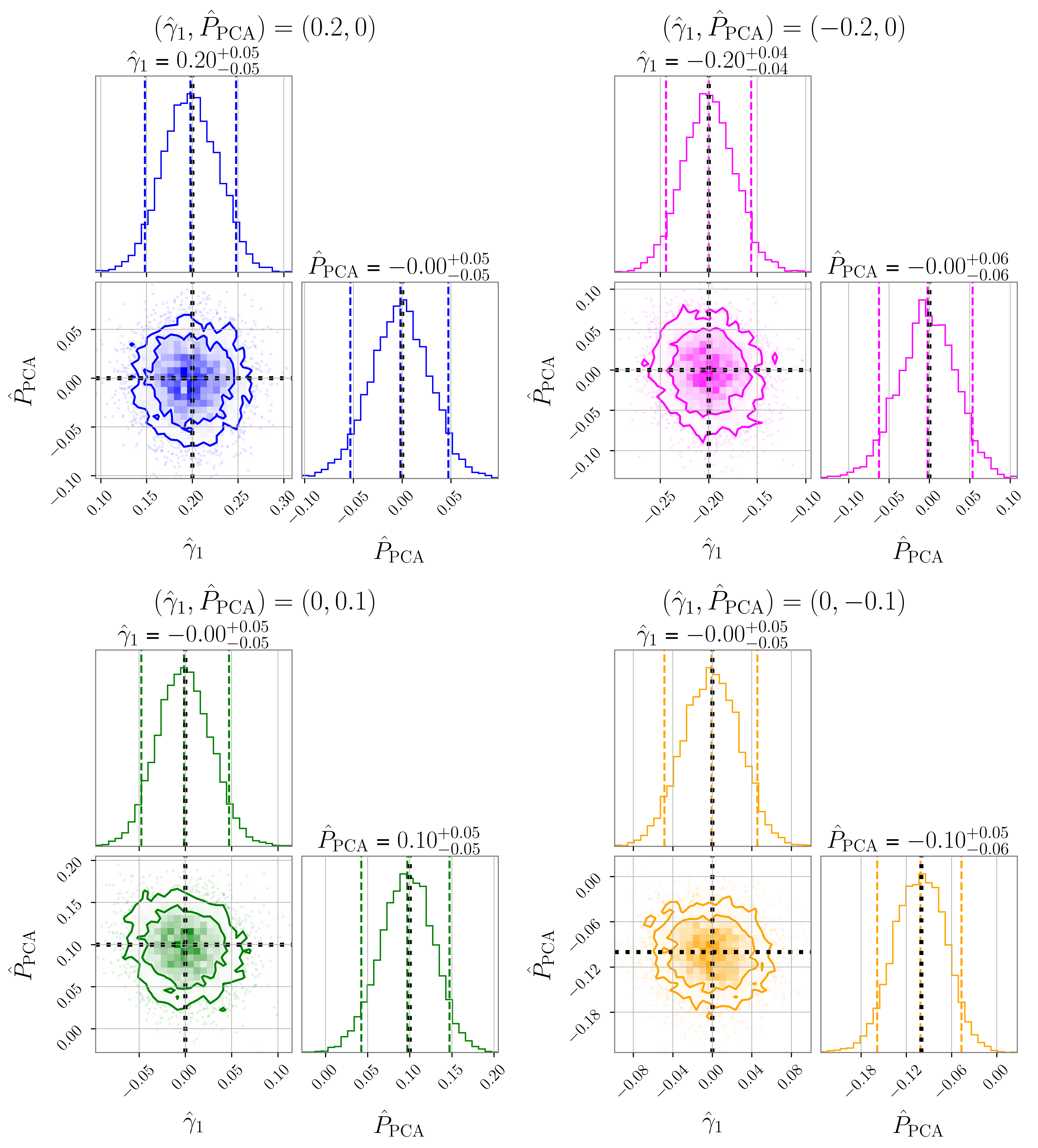} 
    \caption{\Ppds for $\g$ and $\ppca$ with the injection with $(\g, \ppca)=(0.2, 0)$ (top left), $(-0.2, 0)$ (top right), $(0, 0.1)$, (bottom left), and $(0, -0.1)$ (bottom right). The colors are identical to those used in Fig.~\ref{fig:modified_waveform}.}
    \label{fig:inj_60M_beyond-GR_all}
\end{figure}

\begin{figure*}
  \centering
  \begin{minipage}[t]{0.48\textwidth}
    \centering
    \includegraphics[width=\linewidth]{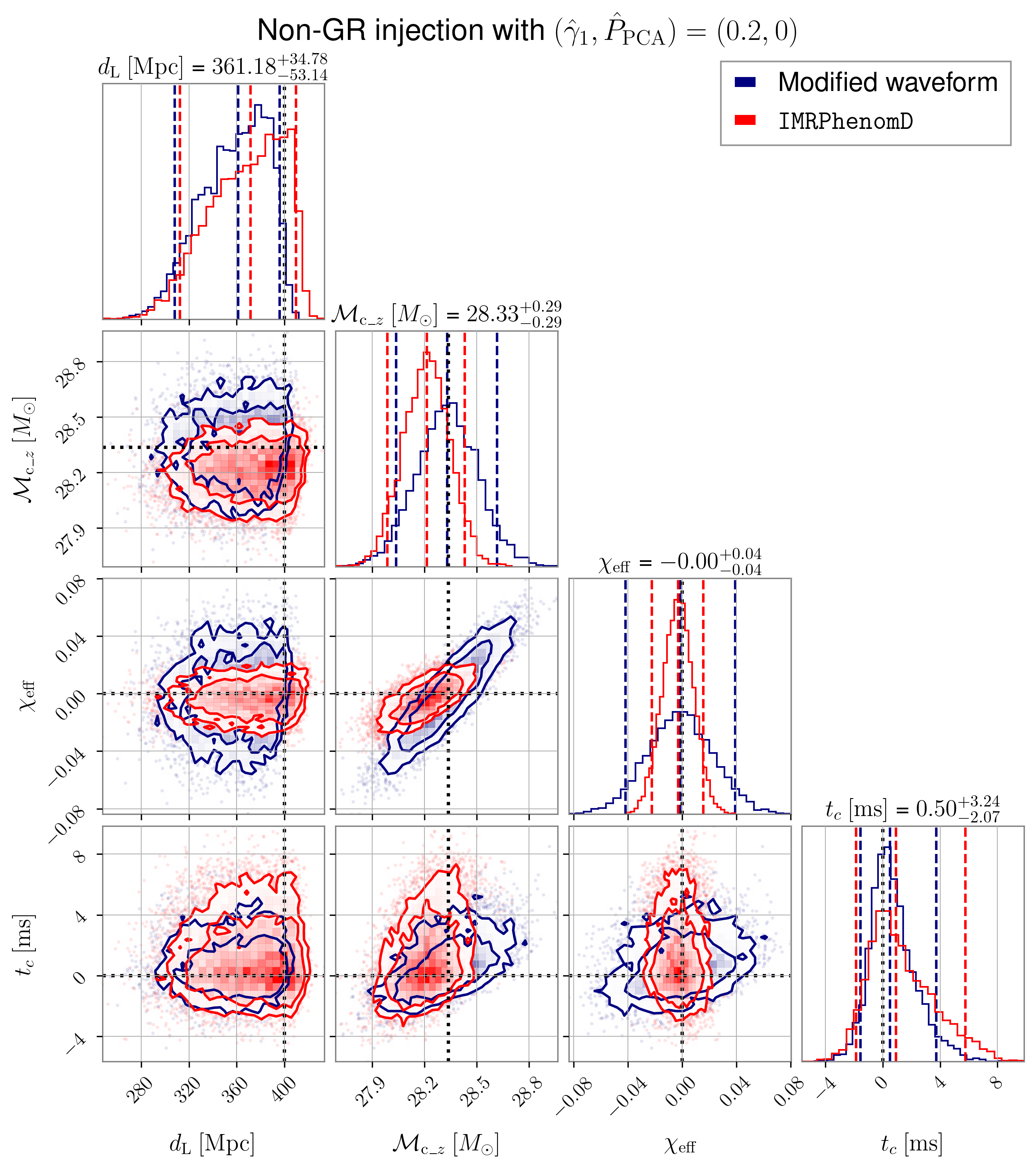}
    \caption{\Ppds for $\{ d_\mathrm{L}, \chi_\mathrm{eff}, t_\mathrm{c} \}$ with the injection with $(\g, \ppca) = ( 0.2, 0)$. The navy and red show the results using the modified and GR waveforms.}
    \label{fig:gamma_inj_standard_params}
  \end{minipage}
  \hfill
  \begin{minipage}[t]{0.48\textwidth}
    \centering
    \includegraphics[width=\linewidth]{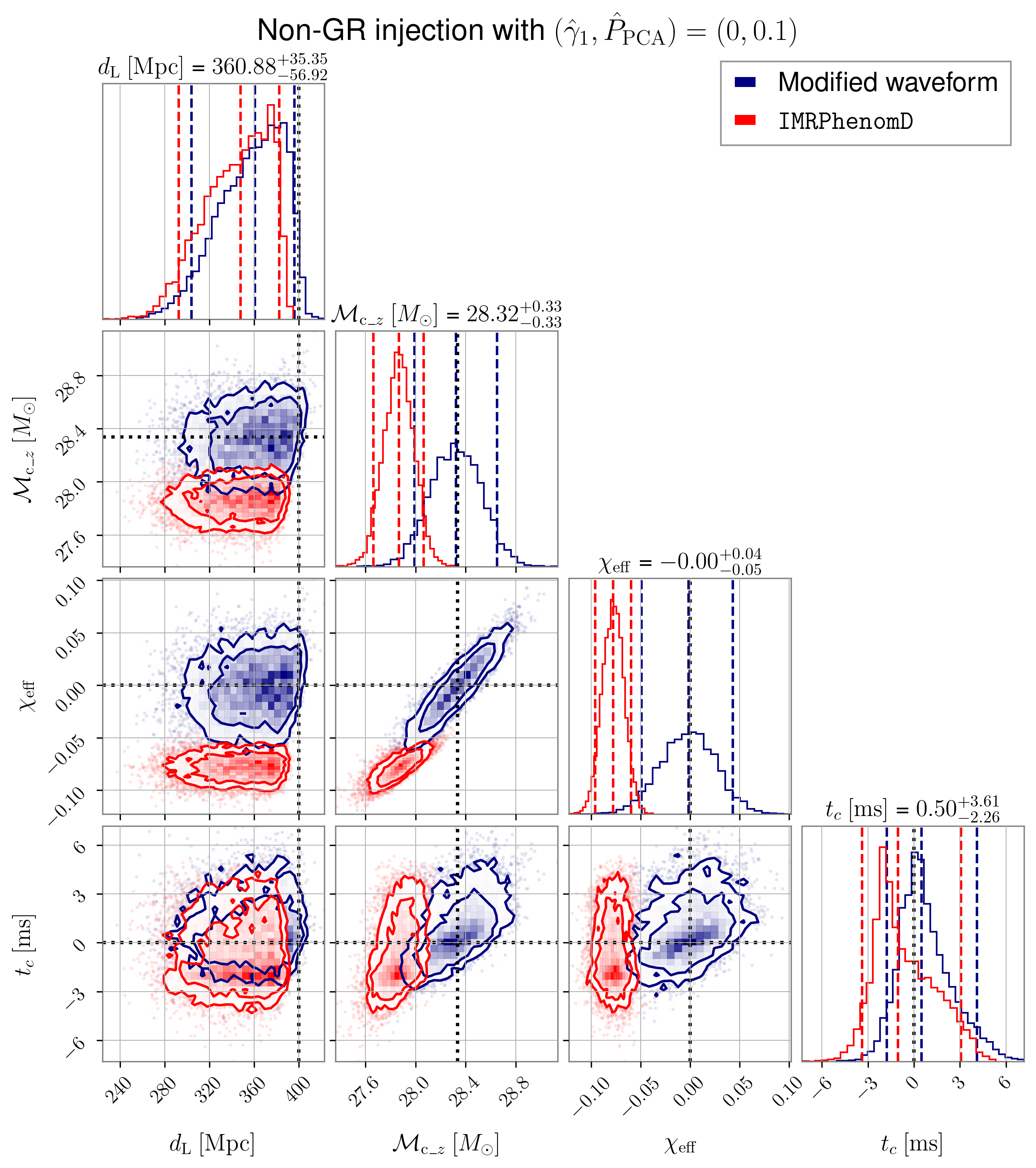}
    \caption{Same as Fig.~\ref{fig:gamma_inj_standard_params}, but the injection with $(\g, \ppca) = ( 0, 0.1)$.}
    \label{fig:P_PCA_inj_standard_params}
  \end{minipage}
\end{figure*}

\subsection{Non-GR injections}
\label{sec:non-GR_injection}
We next perform non-GR injections, varying $\g$ and $\ppca$ as listed in Table~\ref{tb:inj_source_GR}.  
The goal is to test whether the modified waveform can accurately recover the injected deviations.  

In Sec.~\ref{sec:non_GR_inj_beyond-GR}, we focus on the beyond-GR parameters and assess whether the modified waveform can capture their fiducial values. In Sec.~\ref{sec:non_GR_inj_standard_params}, we present the \ppds for the standard GR parameters obtained with both the modified waveform and \texttt{IMRPhenomD}.

\subsubsection{\textbf{Beyond-GR parameters}}
\label{sec:non_GR_inj_beyond-GR} 
Figure~\ref{fig:inj_60M_beyond-GR_all} shows the \ppds for injections with  
$\{ \g, \ppca \} = (0.2, 0)$ (top left), $(-0.2, 0)$ (top right), $(0, 0.1)$ (bottom left), and $(0, -0.1)$ (bottom right).  
The color scheme matches that in Fig.~\ref{fig:modified_waveform}.  
In all cases, the recovered values of $\g$ and $\ppca$ are consistent with the injected values within $68\%$ CIs.  
This indicates that, for the cases investigated here, the modified waveform can correctly recover deviations from GR when they are present in the signal.

\subsubsection{\textbf{GR parameters}}
\label{sec:non_GR_inj_standard_params}
We also analyze the non-GR injection data using \texttt{IMRPhenomD}, to evaluate the impact of neglecting beyond-GR effects on the estimation of $\vec{\theta}_\mathrm{GR}$.  

Figures~\ref{fig:gamma_inj_standard_params} and \ref{fig:P_PCA_inj_standard_params} present the \ppds for  
$\{ d_\mathrm{L}, \mathcal{M}_{\mathrm{c}\_z}, \chi_\mathrm{eff}, t_\mathrm{c} \}$  
for injections with $(\g, \ppca) = (0.2, 0)$ and $(0, 0.1)$, respectively.  
Results from the modified waveform and \texttt{IMRPhenomD} are shown in navy and red. For the injection with an amplitude deviation, both waveforms yield results consistent within $68\%$ CIs. However, for the injection with a phase deviation, the posteriors obtained with \texttt{IMRPhenomD} do not match the injected values,  
even at the 90\% CI.  
This discrepancy arises because some of the GR parameters effectively compensate for the phase deviation through parameter correlations, showing trends similar to those observed in Sec.~\ref{sec:GR_injection}.

Such misidentification of $\vec{\theta}_\mathrm{GR}$ could potentially lead to inconsistencies in the IMR consistency test.  
Some comments for this point are given in Sec.~\ref{subsec:future_work}.

\subsection{Summary}
In this section, we examined the recovery of both GR and non-GR injections using the modified
and GR waveforms. Our main findings are as follows:
\begin{itemize}
    \item \textbf{GR injections:} Introducing the beyond-GR parameters $\g$ and $\ppca$ does not introduce significant systematic biases.  
    The parameter estimates obtained with the modified waveform remain consistent with those from the GR waveform,  
    and the recovered beyond-GR parameters are consistent with zero within $68\%$ CIs.
    \item \textbf{Non-GR injections:} When the signal contains deviations from GR, the modified waveform can reliably recover the injected values of $\g$ and $\ppca$ within $68\%$ CIs.  
    In contrast, the GR waveform can misidentify the binary parameters, as some of the standard GR parameters compensate for the unmodeled deviations, potentially leading to inconsistencies in tests such as the IMR consistency test.
\end{itemize}

These results demonstrate that the modified waveform is robust against false detections of beyond-GR effects when analyzing GR signals, while retaining the ability to accurately capture deviations when they are present. 
This robustness is likely due, at least in part, to the use of PCA in defining $\ppca$: by projecting merger-phase deviations onto their dominant uncorrelated direction.
PCA helps suppress degeneracies with GR parameters, thereby mitigating potential systematic biases.
Encouraged by this robustness, we proceed to apply the modified waveform to real GW data to place direct observational constraints on possible deviations from GR.

\section{Analysis of GW150914}
\label{sec:analysis_of_real_data}
We apply the modified waveform to the first observed GW event, GW150914. 
We focus on GW150914 not only because it has a relatively high network SNR of about 25, but also because it is well consistent with a nearly equal-mass binary without spin precession~\cite{LIGOScientific:2016vlm}, which justifies the assumption that the dominant $(2,2)$ mode provides an accurate description of the signal.
The analysis uses $4~\mathrm{s}$ of strain data centered on the GPS time \texttt{1126259462.4}, following the setup described in Sec.~\ref{sec:injection_recovery}.

In the next subsection, we present the parameter estimation results for $\g$ and $\ppca$, and then translate them into physically interpretable quantities: the additional energy radiated beyond the GR prediction and the deviation in the coalescence time.  
In addition, to demonstrate the applicability of our method to other events, we also present the results for GW190728 in Appendix~\ref{app:GW190728}.

\begin{figure}
    \centering
    \includegraphics[width=\linewidth]{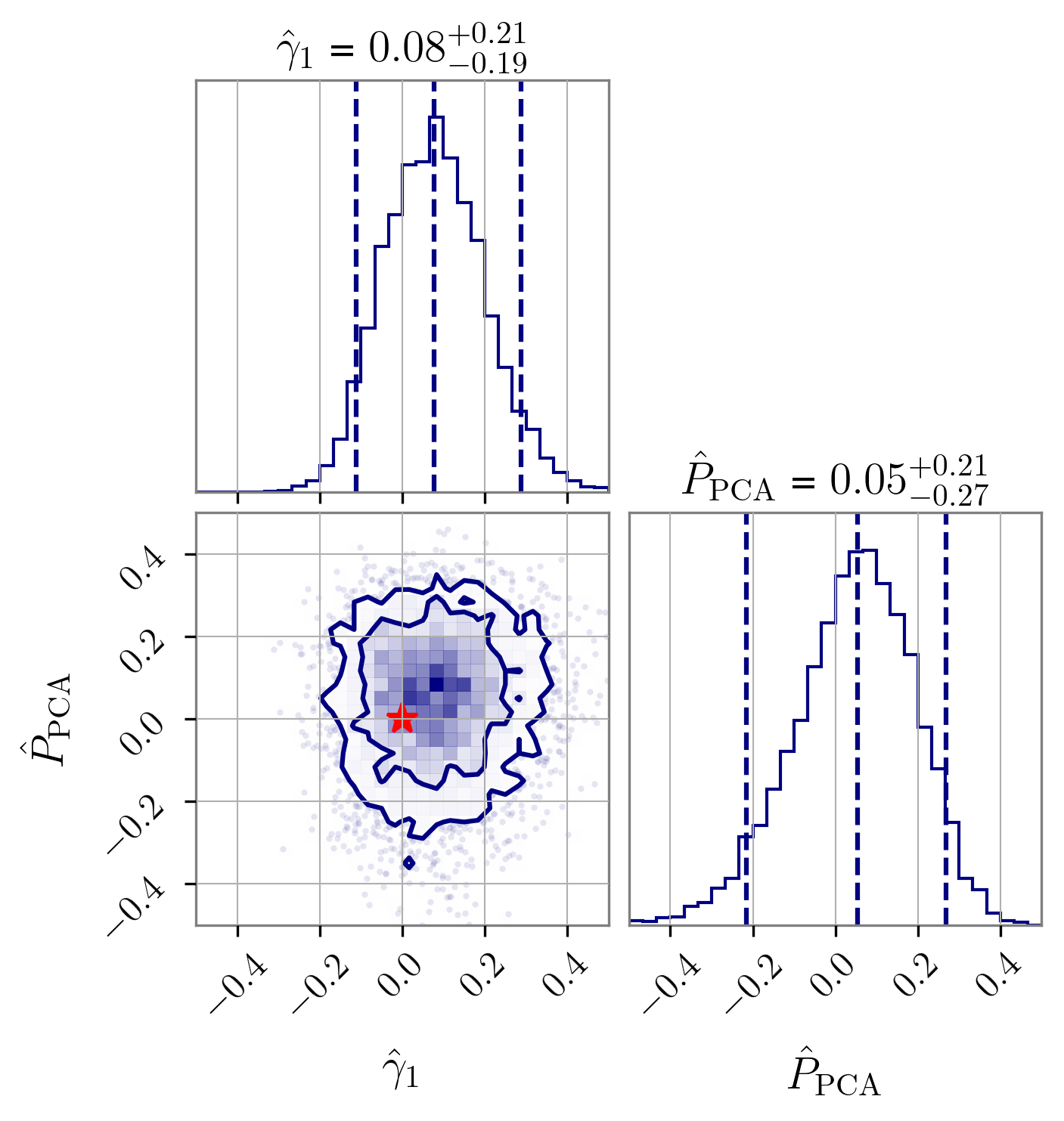}
    \caption{\Ppds for $\g$ and $\ppca$ with GW150914. The dotted lines indicate $90\%$ CI around the mean value. The red star represents $(\g, \ppca)=(0,0)$, which means GR. The data is consistent with GR within 1$\sigma$ error as the contour encloses the star.}
    \label{fig:GW150914_beyond-GR_params}
\end{figure}

\subsection{Results}
\label{sec:beyond-GR_GW150914}
Figure~\ref{fig:GW150914_beyond-GR_params} shows the \ppds for $\g$ and $\ppca$. It indicates that the constraints on the beyond-GR parameters are $\g=0.08^{+0.21}_{-0.19}$ and $\hat{{P}}_\mathrm{PCA}=0.05^{+0.21}_{-0.27}$, both within $90\%$ CIs. The result means that GR is valid within $68\%$ CI since the two-dimensional posterior encloses $(\g, \ppca)=(0,0)$ (red star in Fig.~\ref{fig:GW150914_beyond-GR_params}).

\begin{figure}
    \centering
    \includegraphics[width=\linewidth]{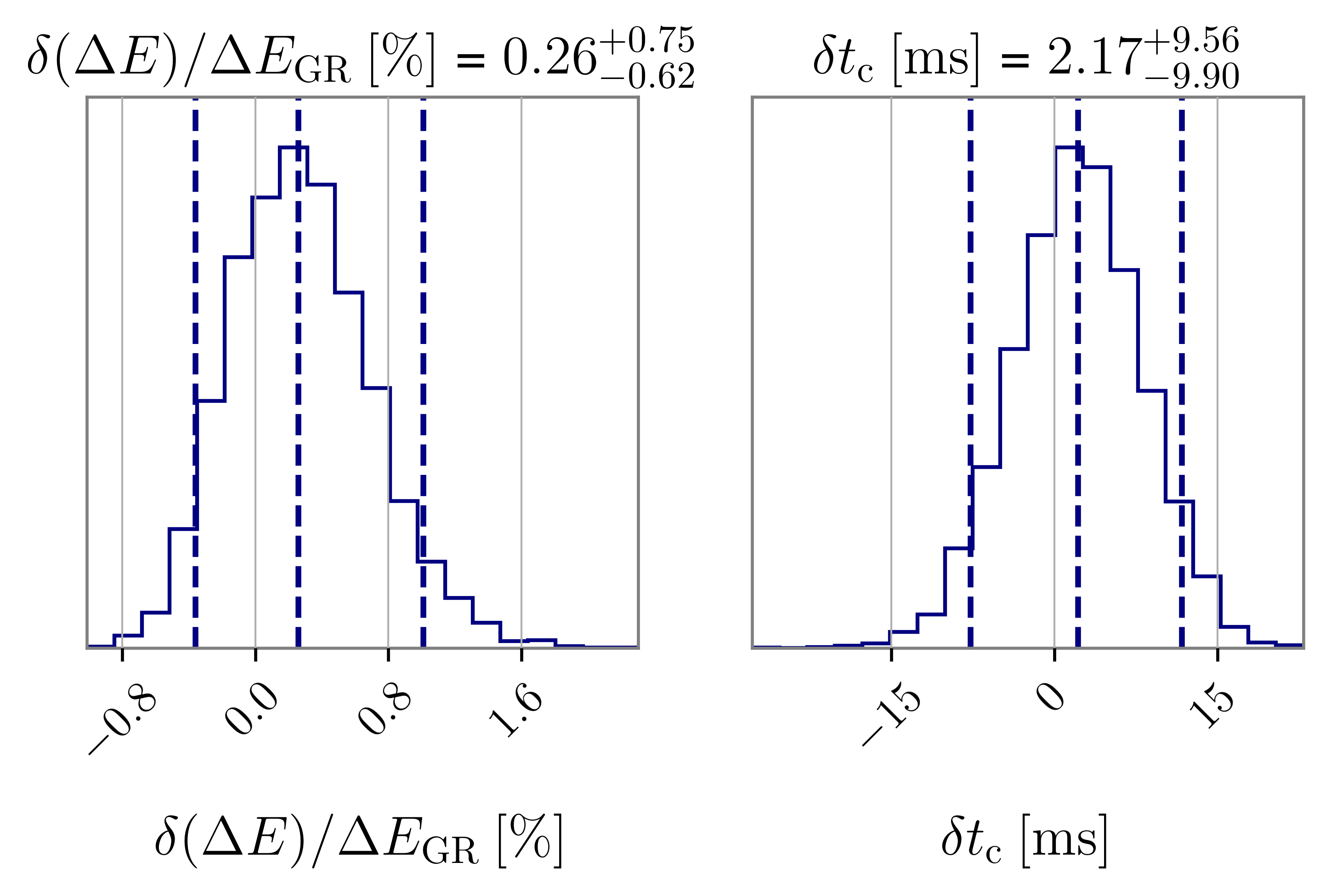}
    \caption{Left: Fraction of the additional energy radiation at the merger stage, $\delta(\Delta E)$, relative to the total GW energy radiation in the IMR process, $\Delta E_\mathrm{GR}$.  
    Right: Deviation in the coalescence time, $\delta t_\mathrm{c}$, induced by $\ppca$, derived using Eq.~\eqref{eq:dev_t_MR}.  
    Note that the right panel does not include the intrinsic $t_\mathrm{c}$ uncertainty of the GR waveform.}
    \label{fig:GW150914_energy_rad}
\end{figure}

The posterior estimates of $\g$ and $\ppca$ can be translated into model-independent constraints on the additional energy radiation and the shift of the coalescence time. Figure~\ref{fig:GW150914_energy_rad} summarizes these results.  

The left panel shows the fraction $\delta(\Delta E)/\Delta E_\mathrm{GR}$, the ratio of the additional energy radiated during the merger stage to the total GW energy radiated throughout the coalescence process predicted by GR. For GW150914, this fraction is constrained to be $0.26^{+0.75}_{-0.62}~[\%]$ within $90\%$ CI, corresponding to a percent-level precision. 

The right panel depicts the deviation in the coalescence time $\delta t_\mathrm{c}$ induced by beyond-GR effects in the merger phase, which is derived based on Eq.~\eqref{eq:dev_t_MR}.  
This expression neglects the reaction of $\g$ on the remnant mass and spin. 
However, as suggested by the left panel, the additional energy loss due to $\g$ is at most at the percent level, so the approximation is expected to be valid within that precision.
For GW150914, the deviation is  
$2.17^{+9.56}_{-9.90}~\mathrm{[ms]}$ within the $90\%$ CI\footnote{We intentionally do not rescale this constraint into units of $M$. 
This choice emphasizes that the result is specific to GW150914, since in scenarios with higher-curvature corrections the deviation would scale with the BH mass~\cite{Payne:2024yhk}, and the bound would not straightforwardly apply to binaries of different total mass.}.
A direct inclusion of $\delta t_\mathrm{c}$ as a beyond-GR parameter and its simultaneous estimation with the GR coalescence time would lead to a complete degeneracy between the two. 
However, by introducing $\ppca$, which corresponds to a principal component of the fractional deviations in the merger-phase coefficients, we can probe the beyond-GR-induced shift in $t_\mathrm{c}$ while breaking the degeneracy with the GR coalescence time.

\section{Discussion}
\label{sec:discussion}
In this section, we first compare our results with previous works focusing on the merger stage. 
We then discuss planned extensions of our work, focusing on two key aspects: (i) establishing a direct comparison with existing testing-GR analyses by the LVK collaboration, and (ii) developing a modified waveform based on the latest phenomenological models to include sub-dominant modes and spin precession. 
These improvements aim to address current limitations and broaden the applicability of our method.

\subsection{Comparison with previous works}

Reference~\cite{Maggio:2022hre} proposed \texttt{pSEOBNRHM}, an extension of the \texttt{SEOBNRHM} model~\cite{Bohe:2016gbl, Cotesta:2018fcv}, designed to parametrize possible deviations from GR in the merger and ringdown phases. 
In the analysis of GW150914, they considered three additional beyond-GR parameters: a deviation parameter $\delta A$ characterizing amplification in the dominant mode in the nonlinear regime, and deviations in the fundamental QNM frequency, namely in the real oscillation frequency $\delta f_{220}$ and the damping time $\delta \tau_{220}$.

Although $\delta A$ and our parameter $\g$ are defined on different baseline GR waveforms, they play a similar role in capturing amplitude deviations in the merger–ringdown regime. 
Our estimate of $\g$ is closely consistent with their constraint, $\delta A = 0.03^{+0.29}_{-0.20}$ (90\% CI), providing a useful cross-check between the two approaches.
Unlike Ref.~\cite{Maggio:2022hre}, however, we do not treat the deviations in the fundamental QNM frequency as independent parameters. 
In our framework, these shifts arise from radiation reaction effects and are therefore functions of $\g$. 
Moreover, while Ref.~\cite{Maggio:2022hre} did not attempt to estimate a possible shift in the coalescence time $\delta t_\mathrm{c}$, owing to degeneracies with standard GR parameters, our method allows such an estimate through the parameter $\ppca$, mitigating correlations with GR parameters.

Finally, their analysis incorporates sub-dominant modes in the waveform, and Ref.~\cite{Pompili:2025cdc} further includes spin-precession effects. 
In contrast, the applicability of our present model is still limited in this respect, which we comment on in the next subsection.

\subsection{Future extensions}
\label{subsec:future_work}

\begin{itemize}
\item \textbf{Comparison with the testing GR analyses by the LVK collaboration}\\
Our method parametrizes and estimates possible deviations in the nonlinear gravity regime, enabling direct constraints on the physical parameters. On the other hand, the LVK collaboration has performed several consistency checks of GR that do not assume a specific parametrization of deviations~\cite{LIGOScientific:2016vlm, LIGOScientific:2019fpa, LIGOScientific:2020tif, LIGOScientific:2021sio}. These complementary approaches provide different, but mutually informative perspectives on potential departures from GR.
Our waveform is more restrictive than the LVK’s generic, non-parametric tests, but in turn it provides direct and physically interpretable constraints on deviations from GR.

Regarding the residuals test, the LVK analysis subtracts the best-fit GR waveform from the observed signal and applies the \texttt{BayesWave} algorithm to determine whether the remaining residuals are statistically consistent with detector noise~\cite{LIGOScientific:2019fpa, LIGOScientific:2020tif, LIGOScientific:2021sio}. 
A significant coherent residual SNR would indicate a mismatch between the data and the GR prediction. 
While our current approach does not directly perform a residual-based check, it would be possible to assess whether the deviations inferred using our waveform are consistent with detectable residual power.

The inspiral–merger–ringdown (IMR) consistency test in the LVK analyses compares the remnant mass and spin inferred independently from the inspiral and post-inspiral portions of the waveform~\cite{LIGOScientific:2016vlm, Ghosh:2016qgn, Ghosh:2017gfp, LIGOScientific:2019fpa, LIGOScientific:2020tif, LIGOScientific:2021sio, Madekar:2024zdj, Shaikh:2024wyn}. 
Discrepancies between these two estimates can indicate potential deviations from GR. 
By injecting our modified waveforms and analyzing them within the IMR consistency test framework, one could quantitatively determine the level of deviation that would be interpreted as a discrepancy in the remnant quantities. 
Conducting such an investigation is left for future work.

\item \textbf{Modified waveform construction based on the latest phenomenological waveforms}\\
In this study, we employ a modified waveform based on \texttt{IMRPhenomD}, which includes only the dominant $(\ell,m)=(2,\pm 2)$ modes and assumes spins aligned with the orbital angular momentum. 
Consequently, contributions from sub-dominant modes such as $(\ell,m)=(3,\pm 3)$ and $(\ell,m)=(2,\pm 1)$, as well as spin-precession effects, are omitted. 
This simplification is generally adequate for nearly equal-mass binaries with moderate SNR, but in high-SNR events, such effects would become important and can potentially mimic signatures of deviations from GR.

Our analysis suggests that introducing beyond-GR parameters via PCA helps mitigate correlations with standard GR parameters and enables the estimation of physically meaningful quantities in a less degenerate manner. 
Based on this insight, future extensions of our method should incorporate sub-dominant modes and spin-precession effects by utilizing the latest phenomenological waveform model, \texttt{IMRPhenomXPHM}~\cite{Pratten:2020ceb}.
Such an extension would also make it possible to reliably analyze high-SNR events where higher harmonics are detected, such as GW250114~\cite{KAGRA:2025oiz, LIGOScientific:2025slb}, with a significant detection of the $(4,4)$ mode, thereby broadening the scope of our framework.

\end{itemize}

\section{Conclusion}
\label{sec:conclusion}
In this work, we derive the observational constraints on possible deviations from GR in the nonlinear regime of BBH coalescences.  
From the analysis of GW150914, no significant deviation from GR was found, with the additional merger-phase energy radiation constrained to $0.26^{+0.75}_{-0.62}~\%$ of the GR prediction and the coalescence time shift constrained to $2.17^{+9.56}_{-9.90}~\mathrm{ms}$, both at the $90\%$ CI.

In Sec.~\ref{sec:modified_waveform}, we reviewed the construction of the modified waveform, which is based on \texttt{IMRPhenomD}. 
The model retains the analytic expressions for the inspiral and ringdown parts of the GR waveform, while introducing controlled deviations in the merger phase and reaction to the remnant mass and spin.

In Sec.~\ref{sec:injection_recovery}, we carried out systematic injection studies. 
For GR injections, the inclusion of beyond-GR parameters was found not to induce false deviations, with posterior distributions consistent with GR predictions. 
For non-GR injections, the modified waveform successfully recovered the injected deviations within statistical uncertainties, whereas a purely GR waveform led to biased estimates of the binary parameters, potentially affecting IMR consistency tests.

In Sec.~\ref{sec:analysis_of_real_data}, we applied the method to the observational data of GW150914. 
The results were consistent with GR within $90\%$ CIs, yielding constraints of $\g=0.08^{+0.21}_{-0.19}$ and $\ppca=0.05^{+0.21}_{-0.27}$. 
These correspond to bounds on the fractional additional energy radiation in the merger phase of $0.26^{+0.75}_{-0.62}~\%$ and a coalescence time deviation of $2.17^{+9.56}_{-9.90}~\mathrm{ms}$, respectively. 
Importantly, we propose these two physically interpretable quantities as observational benchmarks for deviations in the merger phase, providing a reference point for theoretical studies of beyond-GR effects.

This work demonstrates that the modified waveform can robustly constrain deviations from GR without introducing significant biases in the estimation of standard binary parameters. Future extensions will include comparisons with existing testing-GR analyses by the LVK collaboration and the development of modified waveforms based on the latest phenomenological models incorporating subdominant modes and spin precession, thereby enhancing sensitivity to deviations across a wider range of source configurations.

\begin{acknowledgments}
We thank Yanbei Chen, Brian Seymour, and Kent Yagi for useful discussions, and Kazunori Kohri for valuable comments on the manuscript.
D.~W. is supported by JSPS KAKENHI grant No.~23KJ06945. 
A.~N. is supported by JSPS KAKENHI Grants No.~JP23K03408, No.~JP23H00110, and No.~JP23H04893. 
H.~T. is supported by the Hakubi project at Kyoto University and by JSPS KAKENHI Grant No.~JP22K14037.
H.~I. is supported by JST SPRING, Grant No.~JPMJSP2108. 
\end{acknowledgments}

\appendix
\begin{widetext}
\section{Fitting formula}
\label{app:fitting_formula}
For efficient waveform generation in our analysis, we derive fitting formulas for the deviations in the remnant spin and mass, defined as $\delta a_\mathrm{f} := \bar{a}_\mathrm{f} - a_\mathrm{f}$ and $\delta (\Delta M) := (\bar{M}_\mathrm{f} - M_\mathrm{f})/M_\mathrm{int}$. The fitting error is below $3\%$ across most of the parameter space, with slightly larger deviations near $\eta = 0.1$. This level of inaccuracy is not critical for our purposes, as the analysis focuses on nearly equal-mass binaries where the $(2,2)$ mode dominates the signal.

Since the deviation in the dimensionless remnant spin is determined by $\{\eta, \chi_\mathrm{eff}, \g\}$, we model it as
\begin{equation}
\label{eq:fit_af}
    \delta a_\mathrm{f} := \bar{a}_\mathrm{f} - a_\mathrm{f} 
    = f_1(\eta, \chi_\mathrm{eff})\,\g + f_2(\eta, \chi_\mathrm{eff})\,\g^2 \:,
\end{equation}
where
\begin{equation}
\label{eq:f_i}
\begin{split}
    f_i(\eta, \chi_\mathrm{eff}) = &~\hat{p}^i_{00} + \hat{p}^i_{10} \eta \\
    &+ (\chi_\mathrm{eff} - 1)\left(\hat{p}^i_{01} + \hat{p}^i_{11} \eta + \hat{p}^i_{21} \eta^2\right) + (\chi_\mathrm{eff} - 1)^2\left(\hat{p}^i_{02} + \hat{p}^i_{12} \eta + \hat{p}^i_{22} \eta^2\right) + (\chi_\mathrm{eff} - 1)^3\left(\hat{p}^i_{03} + \hat{p}^i_{13} \eta + \hat{p}^i_{23} \eta^2\right)\;,
\end{split}
\end{equation}
with $i = 1, 2$. The coefficients $\{\hat{p}^i_{jk}\}$ are listed in Table~\ref{tab:delta_coeffs}.

We adopt a similar functional form for $\delta (\Delta M)$, but include cubic terms in $\g$ to improve the fitting accuracy:
\begin{equation}
\label{eq:fit_M}
\begin{split}
    \delta (\Delta M)(\eta, \chi_\mathrm{eff}) 
    &:= \frac{\bar{M}_\mathrm{f} - M_\mathrm{f}}{M_\mathrm{int}} = f_3(\eta, \chi_\mathrm{eff})\,\g 
     + f_4(\eta, \chi_\mathrm{eff})\,\g^2 
     + f_5(\eta, \chi_\mathrm{eff})\,\g^3 \:,
\end{split}
\end{equation}
where the functional form of $f_{3,4,5}$ is identical to Eq.~\eqref{eq:f_i}.  
The coefficients $\{\hat{p}^i_{jk}\}$ are again provided in Table~\ref{tab:delta_coeffs}.

\begin{table*}
\centering
\begin{ruledtabular}
\begin{tabular}{lccccc}
 & $f_1$ & $f_2$ & $f_3$ & $f_4$ & $f_5$ \\
\hline
$\hat{p}_{00}$ &  0.0305501  &  0.00341298 & -0.0143104 &  0.00195221 &  0.00310633 \\
$\hat{p}_{10}$ & -0.827374   & -0.251891   &  0.20667   &  0.0208859  & -0.0268509  \\
$\hat{p}_{01}$ & -0.085494   & -0.108239   & -0.0367986 &  0.0485287  &  0.0208047  \\
$\hat{p}_{11}$ &  0.436502   &  1.05727    &  0.542769  & -0.557459   & -0.265124   \\
$\hat{p}_{21}$ & -0.95875    & -2.49765    & -1.71618   &  1.44376    &  0.780273   \\
$\hat{p}_{02}$ & -0.151878   & -0.14713    & -0.0247531 &  0.0577627  &  0.0193892  \\
$\hat{p}_{12}$ &  1.64607    &  1.71045    &  0.336372  & -0.709253   & -0.248971   \\
$\hat{p}_{22}$ & -3.96763    & -4.29235    & -1.20752   &  1.93542    &  0.759615   \\
$\hat{p}_{03}$ & -0.0532096  & -0.0489773  & -0.006192  &  0.018156   &  0.00552157 \\
$\hat{p}_{13}$ &  0.635974   &  0.593763   &  0.0781003 & -0.226727   & -0.0705718  \\
$\hat{p}_{23}$ & -1.60588    & -1.53762    & -0.29018   &  0.630332   &  0.216087   \\
\end{tabular}
\caption{Coefficients $\hat{p}_{jk}^i$ for the functions $f_1, \dots, f_5$ 
defined in Eqs.~\eqref{eq:fit_af} and \eqref{eq:fit_M} derived using \texttt{NonlinearModelFit} in \texttt{Mathematica}.  
The functions $f_1$ and $f_2$ determine the remnant spin deviation $\delta a_\mathrm{f}$, 
while $f_3, f_4,$ and $f_5$ determine the remnant mass deviation $\delta (\Delta M)$.}
\label{tab:delta_coeffs}
\end{ruledtabular}
\end{table*}

\section{Formulas for the shifts of the peak time and phase due to $\ppca$}
\label{app:formulas_shifts}
This section corrects Appendix B of our previous work~\cite{Watarai:2023yky}. 
The expressions provided here do not include the backreaction of $\g$ on the remnant mass and spin. 
This approximation is justified when the additional energy loss associated with $\g$ is small enough.

Here, we show the formulas for the peak time and phase shifts associated with $\hat{P}_\mathrm{PCA}$, denoted as $\Delta t_\mathrm{MR}$ and $\Delta \phi_\mathrm{MR}$:
\begin{align}
    \Delta t_\mathrm{MR} &=
             \Delta t_\mathrm{int} + \frac{M}{2\pi\eta}\Bigl({f_\mathrm{p2}}^{-1}\Delta\beta_2 + {f_\mathrm{p2}}^{-4}\Delta\beta_3- {f_\mathrm{p2}}^{-2}\Delta\alpha_2 - {f_\mathrm{p2}}^{-\frac{1}{4}}\Delta\alpha_3 \Bigr)\;,\label{eq:dev_t_MR}\\
    \Delta \phi_\mathrm{MR}
            &=\Delta \phi_\mathrm{int} - \frac{1}{\eta}\Bigl[ -\{1-\log(f_\mathrm{p2})\}\Delta\beta_2 - \frac{4}{3}{f_\mathrm{p2}}^{-3}\Delta\beta_3+ 2{f_\mathrm{p2}}^{-1}\Delta\alpha_2 - \frac{1}{3}{f_\mathrm{p2}}^\frac{3}{4}\Delta\alpha_3 \Bigr]\;,\label{eq:dev_phi_MR}
\end{align}
where
\begin{align}
    \Delta t_\mathrm{int}
    &= -\frac{M}{2\pi\eta}\left\{ {f_\mathrm{p1}}^{-1}\Delta\beta_2 + {f_\mathrm{p1}}^{-4}\Delta\beta_3 \right\}\;,\\
    \Delta \phi_\mathrm{int}
    &= -\frac{1}{\eta}\left[ \{1-\log(f_\mathrm{p1})\}\Delta\beta_2+\frac{4}{3}{f_\mathrm{p1}}^{-3}\Delta\beta_3 \right]\;.
\end{align}
Once the values of $\ppca$ and the binary parameters are provided, we can derive $\Delta \beta_2, \Delta\beta_3, \Delta\alpha_2$, and $\Delta\alpha_3$. We use Eq.~\eqref{eq:dev_t_MR} to estimate $\delta t_\mathrm{c}$ in Sec.~\ref{sec:analysis_of_real_data}.
\end{widetext}

\section{Full posteriors of GW150914}
\label{app:full_pos}
\begin{figure*}
    \centering
    \includegraphics[width=1.\linewidth]{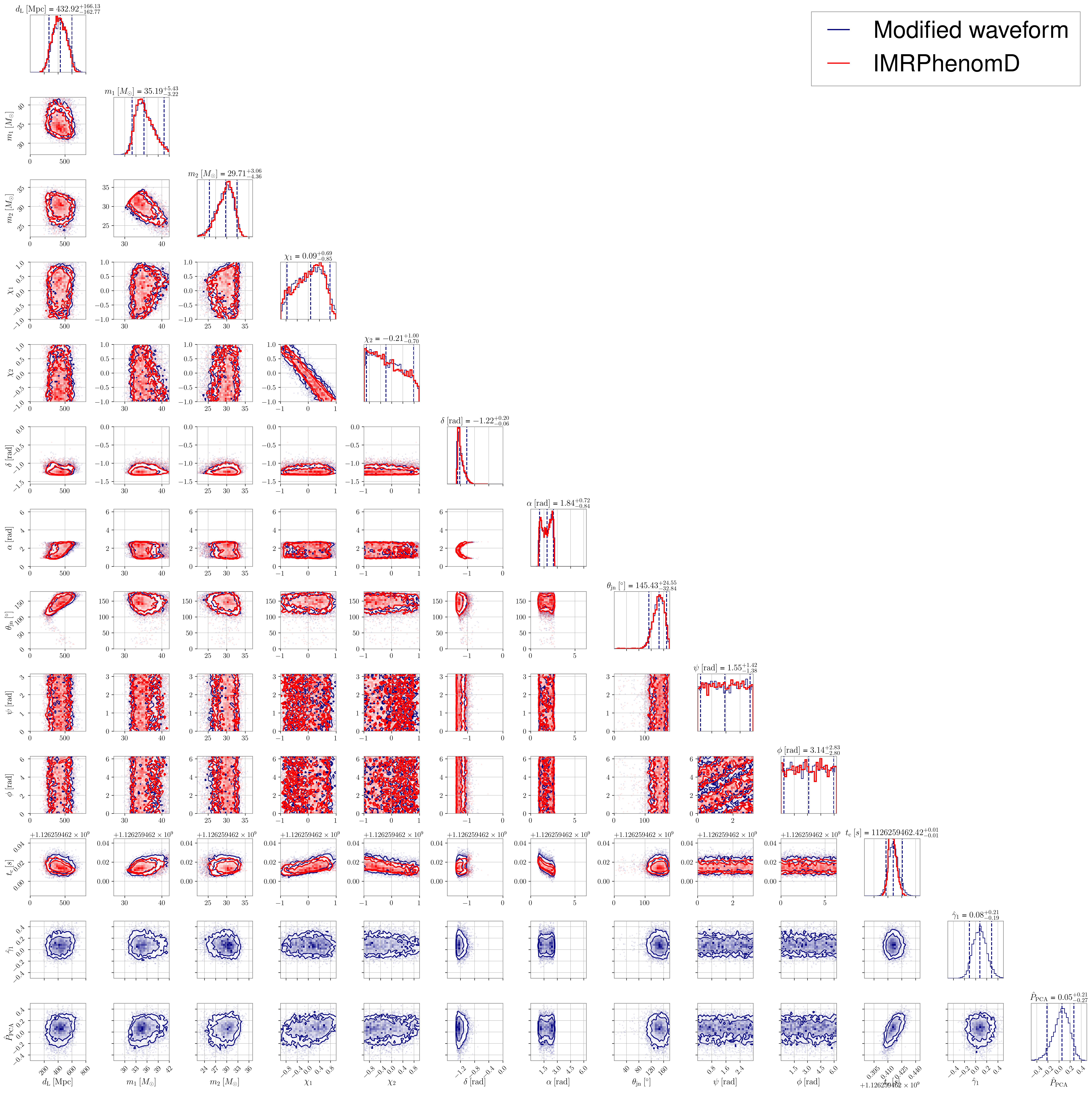}
    \caption{Full \ppds for $\vec{\theta}_\mathrm{GR}$ (red) and $\vec{\theta}_\mathrm{MG}$ (navy) with GW150914. The results for $\vec{\theta}_\mathrm{GR}$ agree with Ref.~\cite{LIGOScientific:2016vlm}. The results for beyond-GR parameters are consistent with zero within $68\%$ CIs. Figure~\ref{fig:GW150914_beyond-GR_params} shows the posterior PDFs only for the beyond-GR parameters.}
    \label{fig:GW150914}
\end{figure*}

Figure~\ref{fig:GW150914} shows the posteriors for all parameters estimated in the analysis. The results for $\vec{\theta}_\mathrm{GR}$ generally agree with one another.

\section{Results of GW190728}
\label{app:GW190728}
As an application of the modified waveform to a binary merger with relatively small component masses,  
we analyze GW190728.  
This event has component masses of $m_1 = 12.3^{+2.7}_{-2.1}~M_\odot$ and $m_2 = 8.1^{+2.0}_{-1.6}~M_\odot$,  
with a source-frame chirp mass of $\mathcal{M} = 8.6^{+0.5}_{-0.3}~M_\odot$ and an effective spin of $\chi_\mathrm{eff} = 0.18^{+0.20}_{-0.23}$~\cite{LIGOScientific:2020ibl}.

We select this event for several reasons.  
First, the mass ratio is close to unity, making the $(2,2)$ mode the dominant contribution.  
Second, the relatively small total mass places GW190728 close to the lighter-mass injection cases investigated in Sec.~\ref{sec:cases}.  
Third, the network SNR is moderately high ($\mathrm{SNR} \approx 13$),  
ensuring sufficient statistical power for parameter estimation.  
Finally, the signal was observed by the three-detector HLV network,  
providing improved sky localization and higher confidence in the astrophysical origin of the event.

\begin{figure}
    \centering
    \includegraphics[width=\linewidth]{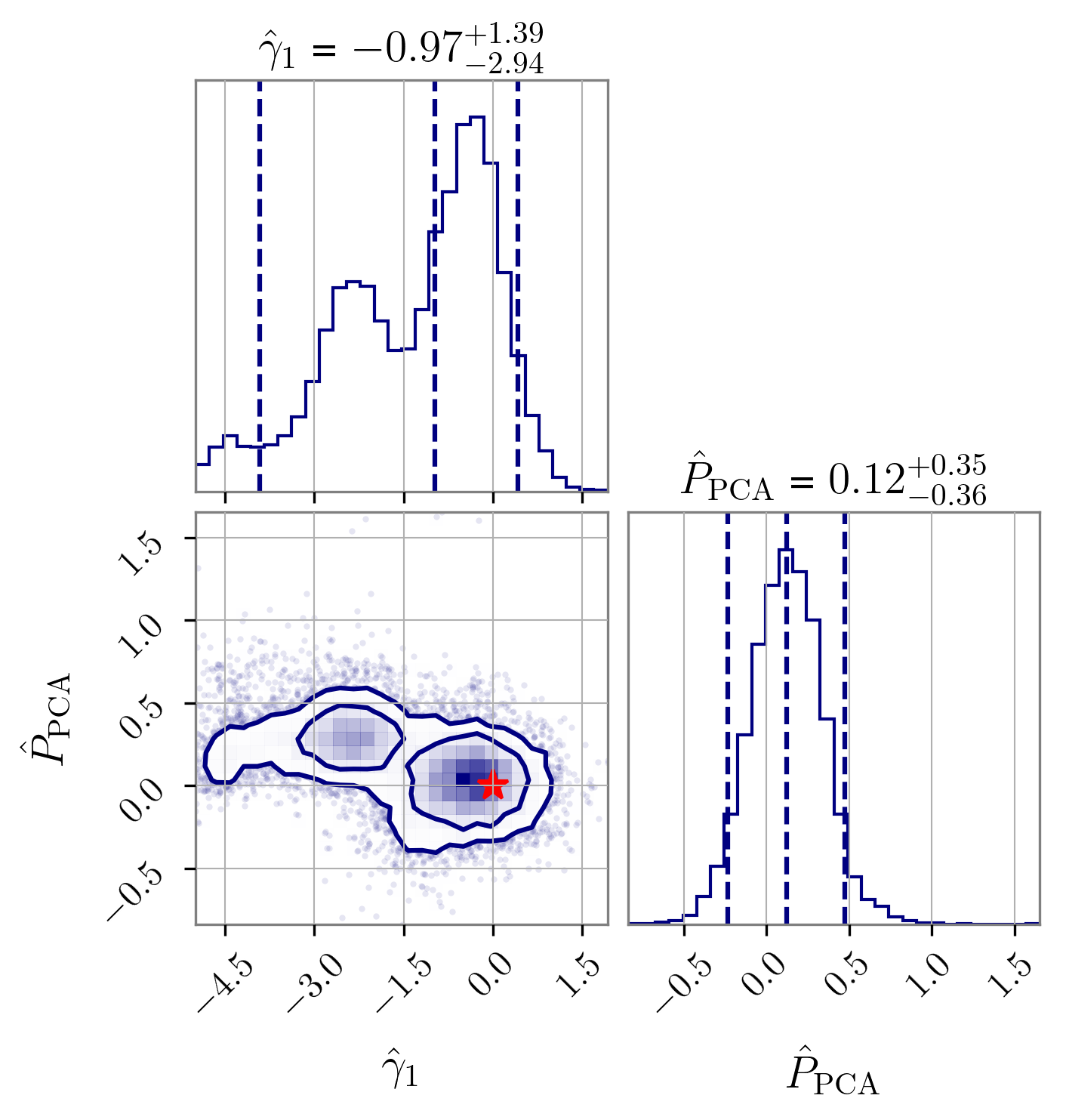}
    \caption{Same as Fig.~\ref{fig:GW150914_beyond-GR_params}, but for GW190728.}
    \label{fig:GW190728}
\end{figure}

Figure~\ref{fig:GW190728} shows the results for $\g$ and $\ppca$.  
The bimodal structure seen in the \ppds for $\g$ arises from the bimodality in the estimation of the GR coalescence time $t_\mathrm{c}$~\cite{pos_GW190728}.  
This feature propagates into $\g$, broadening its statistical uncertainty, while the results remain consistent with GR within the $90\%$ CI.
We do not convert these results into constraints on the additional energy and the shift in the coalescence time, since the inferred values of $\g$ extend beyond its intended validity regime, where the fitting formulas in Sec.~\ref{app:fitting_formula} cease to be accurate. 
As a result, such converted constraints would not be reliable and are therefore not presented here.

\newpage
\bibliographystyle{apsrev4-2}
\bibliography{reference}

\end{document}